\documentclass[onecolumn,prd,amsmath,amssymb,floatfix]{revtex4}

\usepackage{graphicx}
\usepackage{slashed}
\usepackage{amsmath,amssymb}
\usepackage{bm}
\usepackage{epsfig}

\newcommand
{\bpsi}{\bar{\psi}}

\begin{document}

\title{ \bf 
Low energy constituent  quark and pion effective  couplings 
to external electromagnetic  field
and a weak magnetic field 
\footnote{Proceedings of International Workshop on Hadron Physics, Florian\'opolis, BR, 18-23 March, 2018.}}
 
\author{ F\'abio L. Braghin \footnote{ Member of   INCT-FNA,   Proc. No. 464898/2014-5}
\footnote{Support: FAPEG -- 
 Funda\c c\~ao de Apoio \`a Pesquisa do Estado de Goi\'as}
\footnote{email:  braghin@ufg.br}
 }

\affiliation{ Instituto de F\'\i sica, Federal University of Goias,
 Av. Esperan\c ca, s/n,
 74690-900, Goi\^ania, GO, Brazil}

\begin{abstract}
In this work,  a compilation of effective 
 electromagnetic field  couplings  to   
pions and constituent quarks and their effective  interactions
derived previously,
including corrections to  the NJL model, 
 is presented.
The particular case of a weak external magnetic field
along the $\hat{z}$ direction
 is considered  shortly   and 
 effective coupling constants 
are redefined to incorporate the weak-$B_0$ dependence. 
They correspond to corrections
to well known pion-constituent quark couplings   
and 
 to the NJL and vector NJL effective couplings
 that break isospin and chiral symmetries. 
\end{abstract}

\maketitle


\section{Introduction}
\label{intro}

Low energy effective models for hadrons are usually based 
on phenomenology and also general theoretical 
results and symmetries from QCD.
 Nambu-Jona-Lasinio-type  NJL model are  emblematic  so that they are  expected to 
describe  important qualitative effects from QCD such as Dynamical Chiral Symmetry Breaking
DChSB and the emergence of the chiral condensate  
in  the QCD phase diagram.
Effective field theories (EFT)   have been developed and strengthened
and they   contribute for establishing
 these conceptual and calculational gaps between the two
 levels in the description of
strong interactions systems.
 A  large Nc  EFT that copes  the large Nc expansion and 
the constituent quark model was proposed  \cite{derafael}
in Ref. \cite{weinberg-2010}. 
 This EFT is composed by the leading large $N_c$ terms for 
constituent quarks coupled to pions and constituent gluons, 
besides the 
leading terms of chiral perturbation theory.
However, in spite of the phenomenological successes,
 they do not provide microscopic 
first ground numerical predictions for 
 the low energy 
coefficients.
Whereas the light mesons sector have been investigated 
within global color-type models (GCM) and NJL models
in the 1980's and 1990's \cite{ERV,PRC1,NJL},
 the baryon  interactions to mesons
have faced more difficulties. 
The constituent quark model framework assumes 
these baryon effective  interactions are equivalent to the constituent
quark effective interactions.
In 
\cite{EPJA2016,PLB2016,PRD2016,EPJA2018,PRD2018ab}
 we have proposed a QCD  mechanism by which
these baryons-light mesons interactions emerge.
The method will not be explained in the present work
and the reader will find all details in these references.
It considers a Fierz transformation 
from a gluon mediated quark interaction 
to make possible to exploit the whole flavor structure
with  the introduction of
 the light mesons fields by means of
auxiliary field method.
In the present work, a set of resulting interactions are 
shown  mainly for the case when the system undergoes
interactions with a weak background external electromagnetic field.
The recent interest on magnetic field effects on hadron dynamics
\cite{review-B-general}  lead us to show
these interactions
can reduce to those with  a weak magnetic field in a simple way
 when considering only the first Landau orbit.
The resulting terms can be calculated perturbatively for increasing 
strength of the magnetic field as discussed in \cite{weak-B}.

 All the effective couplings presented below are derived 
  from the following    leading term  of  QCD effective action: 
\begin{eqnarray} \label{Seff}  
Z = N \int {\cal D}[\bpsi, \psi]
e^{ i \int_x  \left[
\bar{\psi} \left( i \slashed{D} 
- m \right) \psi 
-
 \frac{g^2}{2}\int_y j_{\mu}^b (x) 
{\tilde{R}}^{\mu \nu}_{bc} (x-y) j_{\nu}^{c} (y) 
+ \bpsi J + J
^* \psi \right] } ,
\end{eqnarray}
where
 the color  quark current is 
$j^{\mu}_a = \bar{\psi} \lambda_a \gamma^{\mu} \psi$, 
and:
$D_{\mu} = \partial_\mu  - i e Q  A_{\mu}$
with  the diagonal matrix 
$\hat{Q} =  diag(2/3, -1/3)$.
In several  gauges the gluon kernel
is  written in terms of
$R_T(k)$, $R_L(k)$ 
In particular it will be assumed and required
 that this dressed  gluon propagator provides enough strength for 
generating 
DChSB, so that a chiral  condensate and the corresponding large effective constituent quark
mass appear.
This can only  be achieved by incorporating  to some extent
the non Abelian gluon dynamics.

\section{ Light mesons and constituent quarks}

  The sea quark determinant is obtained
in terms of the light mesons fields and constituent quark currents.
By neglecting   light vector mesons  that 
are considerably heavier  for  the low energy 
regime, and by performing
a chiral rotation that  eliminates the scalar degree of freedom, 
the determinant  is given by:
\begin{eqnarray} \label{Seff-det}  
S_{eff}   &=&  i \;  Tr  \; \ln \; \left\{
i \left( S_{c}^{-1} (x-y) 
+
\sum_q  a_q \Gamma_q j_q (x,y) \right)
 \right\} 
,
\end{eqnarray}
where 
$Tr$ stands for traces of all discrete internal indices 
and integration of  spacetime coordinates
and  the
 quark kernel can be written as 
\begin{eqnarray}
S_{c}^{-1} (x-y) =  
S_{0,c}^{-1} (x-y) + \Xi  (x-y),
\end{eqnarray}
where
 $S_{0,c}^{-1} (x-y) = \left( i \slashed{D} -  M^* 
\right) \delta (x-y)$ and
 $M^*$ is   the resulting 
 effective  quark mass  from 
the gap equation for the DChSB solution, $M^* = m + < S >$.
The following quantity was used:
$$
\Xi (x,y)    = F ( P_R U + P_L U^\dagger)  \delta(x-y) = F ( P_R e^{i \vec{\pi} \cdot \sigma}
+ 
P_L  e^{ - i \vec{\pi} \cdot \sigma} )  \delta(x-y)
,$$
where $P_{R/L}$ are the chiral right/left hand projectors.
The constituent quark degrees of freedom appear in terms of 
 quark  flavor currents $  j_q (x,y)$, 
and with the Pauli matrices for SU(2) isospin and 
 Dirac matrices can be written as:
\begin{eqnarray} \label{Rq-j}
\frac{\sum_q  a_q \Gamma_q  j_q (x,y)}{ \alpha g^2}
&=&
2   R (x-y)
 \left[  \bpsi (y) \psi(x)
+ i  \gamma_5 \sigma_i  \bpsi (y) i \gamma_5  \sigma_i \psi (x)
+    \bpsi (y) \sigma_i \psi(x)
+ i  \gamma_5  \bpsi (y) i \gamma_5  \psi (x)
\right]
\\
&-& 
 \bar{R}^{\mu\nu} (x-y) \left\{ \gamma_\mu  \sigma_i \left[
 \bpsi (y) \gamma_\nu  \sigma_i \psi(x)
+  i \gamma_5   \bpsi  (y)
i \gamma_5 \gamma_\nu  \sigma_i \psi (x) \right]
- \gamma_\mu   \left[
 \bpsi (y) \gamma_\nu  \psi(x)
+  i \gamma_5   \bpsi  (y)
i \gamma_5 \gamma_\nu   \psi (x) \right] \right\} .
\nonumber
\end{eqnarray}
Saddle point  equations are calculated  for each of the auxiliary  field
and only the scalar auxiliary field develops a classical counterpart
by neglecting the eventual magnetic field.

\section{First order constituent photon-quark-pion effective couplings}

From a very quark and gluon effective mass expansion 
the leading effective constituent quark-pion  terms
and their leading (dipolar type) couplings with the external photon field
   arise in the large quark mass and 
local very longwavelength  limit:
\begin{eqnarray} \label{pion-q-couplings}
{\cal L}_{Q\pi} &=&
g_{2} F \;  tr_F \; Z_+  j_s + g_{1} F  \; tr_F (\sigma_i Z_-) j_{ps}^i
+  i 2 g_{V}  \;  tr_F ( \sigma_i \partial_\mu Z_+ )  j^{V,\mu}_i 
+
 i 2  g_{A}   \;  tr_F ( \sigma_i  \partial_\mu Z_- ) j_{A,\mu}^i ,
\\
 \label{q-pi-A}
{\cal L}_{Q\pi A} &=&
M_{FF} F_{\mu\nu}F^{\mu\nu} \;  j_s 
+
 g_{vmd}
A_\mu \; 
j^\mu_{i=3}
+ 
g_{F-js-\pi} \; F  \;  F_{\mu\nu}^2
  tr_F  ( \left\{ Q ,  Z_+  \right\} Q )
 j_s
\\
&+&  i 
g_{F-ps-\pi} \; F  \;  F_{\mu\nu}^2
 tr_F \left(
 \left[ Q, Z_- \right]  \left[ Q ,  \sigma_i \right]
\right)
 j_{ps}^i
+   i g_{jV A } \; F \; 
 F^{\mu\nu} tr_F ( \left\{ \partial_\mu Z_+ , Q  \right\}
\sigma_i  + \left[ Q ,\ \sigma_i \right] \partial_\mu Z_+ 
 ) 
j_{V,\nu} ^i
\nonumber
\\
&+& i  g_{jA A } \; F \; 
 F^{\mu\nu} tr_F ( [ Q ,  \partial_\mu Z_- ]  \sigma_i 
+ \left\{ Q , \sigma_i \right\} \partial_\mu Z_- )
j_{A,\nu} ^i 
\nonumber
\; + \; {\cal O}(A_\mu^2),
\end{eqnarray}
where 
$Z_\pm = \frac{1}{2} (U \pm U^\dagger)$.
In the expression (\ref{pion-q-couplings})
  $g_{1}, g_{2}$  correspond respectively to the usual 
pseudoscalar and    scalar couplings of one and two pions to a pseudoscalar/scalar quark currents.
 $g_{V}$  the two pion coupling
  to a vector quark current
 and   $g_A$ the usual axial coupling.
It has been found $g_V=g_A$ at this level and the Goldberger Treiman relation
\cite{GT-ori}
is satisfied \cite{EPJA2018}.
In all these expressions $tr_F$ stands for the trace in isospin indices.
In expression (\ref{q-pi-A}) the leading couplings to an external photon field
are shown.
The 
canonical normalization of the pion field
requires a multiplicative  factors $1/F$  
to  redefine   coupling constants.
The expressons for the effective coupling constants 
 in terms of the components of
quark and gluon kernels for the case of zero external magnetic field 
can be found in \cite{EPJA2016,EPJA2018} and they
are not presented here.

 If the above external photon correspond to  a weak external magnetic field
$ A^\mu  = B_0/2 ( 0, - y,  x, 0)$ the above expression (\ref{q-pi-A}) except the VMD
(vector meson dominance)
term,  reduce
to the following weak  magnetic field dependent  expression:  
\begin{eqnarray} \label{q-pi-B}
{\cal L}_{q\pi-A} &=&
  \bar{M}_{B}   j_s 
+
\bar{g}_{Fjs\pi}   \;
  tr_F  ( \left\{ Q ,  Z_+  \right\} Q )
 j_s
+  i 
\bar{g}_{Fps\pi}  \; 
 tr_F \left(
 \left\{ Q, Z_- \right\} \left\{ Q ,  \sigma_i \right\} 
\right)
 j_{ps}^i
\nonumber
\\
&+& i 
\bar{g}_{jV A }   tr_F ( \left[ \partial_x Z_+ , Q  \right]
\sigma_i 
 ) 
j_{V, y}^i
+  i  \bar{g}_{jA A }  tr_F ( [ Q ,  \partial_x  Z_- ]  \sigma_i ) 
j_{A,y} ^i 
,
\end{eqnarray}
where the expressions for the effective coupling constants 
$\bar{g}$
by accounting 
the leading Landau orbit were given in \cite{EPJA2018}
and the stronger magnetic field case will be analysed elsehwere.

\begin{figure}[ht!]
\centering
\includegraphics[width=120mm]{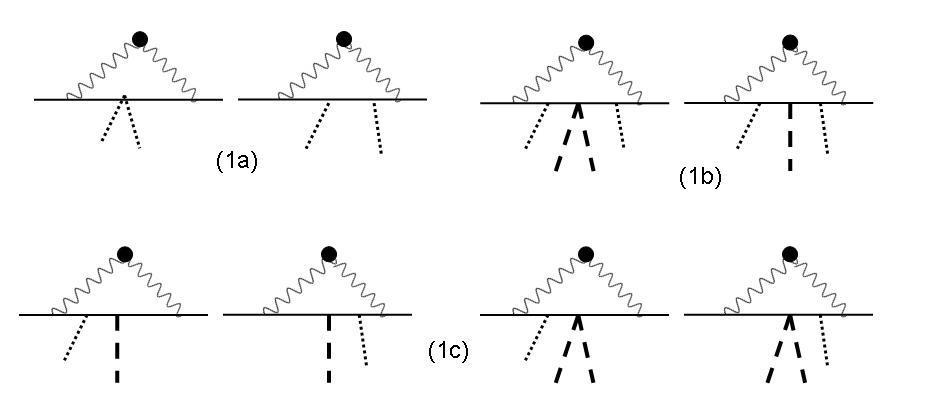}
\caption{ \label{diagrams-2}
\small
In these diagrams, the wavy line
 with a full circle is a (dressed) non perturbative gluon propagator,
a  dotted line represents the external photon field tensor $F^{\mu\nu}$,
 whereas 
the dashed line stands for the pion.
Diagrams (1a)
two contributions $\Delta_B M^*$ for the constituent quark effective mass
due to the electromagnetic coupling.
Diagrams (1b) correspond respectively to electromagnetic coupling of
 the  scalar  and pseudoscalar pion couplings to 
constituent quark.
Diagrams (1c) the electromagnetic coupling of  the 
axial and vector  pion couplings to the constituent quark..}
\end{figure}

The  traces in flavor  indices of the Pauli matrices with 
the matrix $Q$ given after expression (\ref{Seff}) were computed
for  the leading weak pion field: 
$U \simeq 1 + i \vec{\tau} \cdot \vec{\pi} 
+ ...$ and
$ U^\dagger \simeq 1 - i \vec{\tau} \cdot \vec{\pi} 
+
 ...$.
The leading terms of the
effective couplings above
 can be written 
as:
\begin{eqnarray}
{\cal L}^{q-\pi}_{U,B} &=&
\left(
g_2 
-
 \bar{g}_{\pi-B_0}    
  \frac{5}{9}
\right)
 \vec{\pi}^2  
 j_s
+ 
\left( g_1  \delta_{ij} -
\bar{g}_{\pi jB_0} \; 
 \frac{4}{3} \epsilon_{ij3}
\right) \pi_i
 j_{ps}^j  ,
\\
&+& \left( g_V \epsilon_{ijk}-   \bar{g}_{jV A B_0}
T_{jki}  \frac{4}{3}\right)
 \pi_j 
(\partial_\mu \pi_k)
j_{V,\nu}^i
+ \left( g_A \delta_{ij} -    \bar{g}_{jA B_0}   \frac{4}{3} \epsilon_{ij3}
\right)  \partial_\mu \pi_i \; 
j_{A,\nu} ^j ,
\end{eqnarray}
where 
$T_{jki} = 
\delta_{ij}\delta_{3k} - \delta_{j3} \delta_{ik}$.
This form makes clear that  the electromagnetic coupling 
is different for the charged pions  and constituent quarks,
and,  besides that, it shows more than one channel of the 
corresponding interactions. 
The corresponding couplings for 
light vector and axial mesons were presented in 
\cite{PRD2018ab}.

\section{ Corrections to the NJL model}

By neglecting the mesons fields, the leading large quark and gluon effective  masses
 terms for 
constituent quark self intearctions  from the expansion of the determinant
within  the very long-wavelength (local) limit  are given by:
\begin{eqnarray} \label{4quark}
{\cal L}_{4q} 
&=&  
\; 
\Delta_B M^* \bpsi \psi + 
  g_{4,B}  \left[ ( \bpsi  \psi )^2
 +   (  \bpsi \sigma_i  i \gamma_5 \psi )^2  \right]
+
G_1^{ij}
 (  \bpsi \sigma_i  i \gamma_5 \psi )
 (  \bpsi \sigma_j  i \gamma_5 \psi )
\nonumber
\\
&+&
\left[  \delta_{ij} G_2
 + i  \epsilon_{ij3} 
G_3^{ij}
\right]
\left[ ( \bpsi  \sigma_i \gamma_\mu \psi ) ( \bpsi  \sigma_j \gamma^\mu \psi )
 + ( \bpsi \sigma_i \gamma_\mu \gamma_5 \psi )  
( \bpsi \sigma_j \gamma^\mu \gamma_5 \psi )
\right] 
\nonumber
\\
&+&
g_{s,sb}   ( \bpsi  \psi )^2 
+ 
g_{v,sb} ( \bpsi  \sigma_i \gamma_\mu \psi )^2
+ g_{vmd} A_\mu \bpsi \gamma^\mu \sigma^3 \psi
,
\end{eqnarray}
where
\begin{eqnarray}
G_1^{ij} &=&
\left(  \frac{3 \bar{g}_{4,B}}{5} i \; \epsilon_{ij3}
+  \bar{g}_{ps,B} \;c_i \; \delta_{ij} \right),
\nonumber
\\
G_2 &=&  \left( \bar{g}_{4v,B}  + \bar{g}_{4v,B2}  \;c_i \;  
+ \frac{\bar{g}_{4v,B-F}}{3}  
+ \bar{g}_{4v2,B}
\right)
\\
G_3^{ij} &=&
\left( \frac{3}{5} \bar{g}_{4v,B} 
+ \bar{g}_{4v,B-F} 
+ 3 \bar{g}_{4v2,B}
\right)
\end{eqnarray}
and where the coupling constants 
$g_{4,B}, g_{ps,B}, g_{4v, B}, g_{4v,B2}, g_{4v, B-F}, g_{4v2,B}$
are magnetic field dependent coupling constants expressed as 
functions of components of  the quark and gluon kernels given in \cite{PRD2016}.
The effective couplings with  $g_{s,sb},g_{v,sb}$ are chiral symmetry breaking 
ones  and they were discussed in \cite{PLB2016}.
The last term represents the photon coupling to a neutral vector meson rho 
coupling \cite{PRD2018ab}.
The following notation was adopted in the terms depending on the coefficients $c_i$
with operators $\Gamma_i$: 
$ \;c_i \; 
 (  \bpsi \Gamma_i \psi )^2  
=  \;c_1 \; 
 (  \bpsi  \Gamma_1 \psi )^2  
+ \;c_2 \; 
 (  \bpsi  \Gamma_2 \psi )^2 
+ 
\;c_3 \; 
 (  \bpsi  \Gamma_3 \psi )^2 $,
being defined 
the following isospin   coefficients:
$c_1=-\frac{4}{9}$,
$c_2 = \frac{4}{9}$ and
$c_3 = \frac{5}{9}$.
These expressions also make clear the different electromagnetic 
couplings of currents of  charge quarks, although it also presents different channels 
of the interaction in which neutral quark currents also appear.

 \begin{table*}[ht]
\caption{
\small 
 In the first column the following set of 
 values    are displayed 
$M^*$ for given  $\Lambda_i$ and $h_a$,
being that $h_a$ is a factor representing 
 quark gluon coupling constant .
This factor was chosen to reproduce the value of the pion vector or axial 
coupling constant $g_{ v} h_a = 1$ and it multiplies the gluon propagator.
The gluon propagator taken from
Ref. \cite{cornwall}.
From the second  to the last  columns, values for some 
of the effective coupling constants and parameters
from  the expressions
presented in \cite{EPJA2018}
for the  usual  pion field definition in terms of the functions
$U,U^\dagger$.
The last two columns show two of the quark-quark effective coupling constant
correction due to a very weak magnetic field divided by 
an estimate for the $g_4$ NJL model coupling constant
obtained  from the same method.
 (e.v.) in the last line stands for  some experimental or expected values. 
In this, it was assumed the constituent quark mass of pion should be
half of the pion mass $140$MeV
and the constituent quark mass one third of the nucleon mass $939$MeV.
The larger values of $M^*$ are obtained as consequence of the 
magnetic catalysis due to $B_0$.} 
\centering 
\begin{tabular}{c c c 
c c c c c c c c } 
\hline\hline 
& $M^*$ \; $h_a$ \; $\Lambda$
& $M_3 h_a$ & $g_{ps} h_a$ & $g_v h_a$ & 
$\frac{\bar{M}_B h_a}{(\frac{eB_0}{{M^*}^2})^2}$ & 
$\frac{\bar{g}_{ps}^B h_a}{(\frac{eB_0}{{M^*}^2})^2}$ & 
$\frac{\bar{g}_V^B h_a}{(\frac{eB_0}{{M^*}^2})}$ & $g_{vmd} h_a$
& $\frac{g_{4B} }{g_{4}} \frac{1}{\frac{(e B_0)^2}{{M^*}^4}}$ & 
$\frac{g_{4F-B}}{g_{4}} \frac{1}{\frac{(e B_0)}{{M^*}^2}}$
\\
& (GeV)
\;  -  \; (GeV) \; 
& (MeV)  & -& - & (MeV) &  - &  - & (MeV)$^2$ & -
&  -
\\ 
\hline
\\ [0.5ex]
& 0.45 \; $\frac{1}{0.16}$ \; 0.600  &
2760 &  0.9  &   1 &  1556  &  4.4  & 1.1  & 3.0 & $ 1.0 \; 10^{-4}$& 
$1.4 \; 10^{-4}$
\\ [0.5ex]
& 0.41 \; $\frac{1}{0.18}$ \; 0.600  &
2672
 & 1   &   1 & 1289  &  3.9   &   1.2 & 2.9 & $1.0  \; 10^{-4}$ & 
$1.3 \; 10^{-4}$
\\ [0.5ex]
& 0.30 \; $\frac{1}{0.30}$ \; 0.575 &
1752 & 2.6    &   1 &  628  &  1.8  & 1.1  & 0.2 &
$1.1 \; . 10^{-4}$ &  $1.5 \; 10^{-4}$
\\ [0.5ex]
 & 0.07  \; $\frac{1}{0.76}$ \; 0.450 &   
335  &  9.3  & 1  &  18     &  0.2   &  0.6  & 0.2 &
$- 2.0 \;  10^{-5}$ & $1.0 \; 10^{-6}$
 \\ [0.5ex]
e.v.&  0.07 \; \; 1  \;  \; \; -  \; \;  &    313 
&  13.5  & 1 &  -   &  -   &  - & - & - & - 
\\[1ex] 
\hline 
\end{tabular}
\label{table:results-2} 
\end{table*}

\section{Summary and final remarks}

 A collection of 
the leading electromagnetic couplings  to 
pions and constituent 
quarks was presented 
extracted from \cite{EPJA2018,PRD2016}.
  A large effective quark mass expansion for the sea quark determinant 
yielded 
  different known  pion effective couplings to quarks:
vector, axial, pseudoscalar and scalar.
The corresponding couplings to the electromagnetic field explicitely
break  chiral and isospin symmetries,
and they have been considered for
 $\frac{(eB_0)}{{M^*}^2} << 1$.
It is interesting to note that, in the leading order terms, 
 the weak magnetic field does  not
mix the contribution of each of the gluon propagator components,
transversal or longitudinal,
what has been shown by considering only the 
 leading contribution  from the leading Landau orbit
according to \cite{weak-B}.
It is remarkable that the best agreement for the known effective 
parameters in the Table were obtained for 
a quark effective mass $70$MeV that is half of the pion mass.
This is the effective mass obtained from the gap equation
of the type of the GCM or NJL models, and 
it was associated to sea quarks.
The constituent quark effective mass  was associated rather 
to $M_3$ that is an effective parameter in the resulting effective model.
A more complete account of the Landau orbits for the stronger 
magnetic field cases will be presented elsewhere.

{\bf Acknowledgments}
The author thanks  short discussions with 
 G.I. Krein, C.D. Roberts,  I.A. Shovkovy 
and J. O. Andersen.
The author participates of the 
project   INCT-FNA,   Proc. No. 464898/2014-5
and he thanks  Funda\c c\~ao de Amparo \`a Pesquisa do 
Estado de Goi\'as- FAPEG for financial support.

\end{document}